# Time delayed laser networks: phase versus chaos synchronization

I. Reidler[1], M. Nixon[2], Y. Aviad[1], S. Guberman[3], A. A. Friesem[2], M. Rosenbluh[1], N. Davidson[2] and I. Kanter[3,4,*]

[1]Department of Physics, The Jack and Pearl Resnick Institute for Advanced Technology, Bar-Ilan University, Ramat-Gan, 52900 Israel

[2]Department of Physics of Complex Systems, Weizmann Institute of Science, Rehovot76100, Israel

[3]Departmet of Physics, Bar-Ilan University, Ramat-Gan52900, Israel

[4]Gonda Interdisciplinary Brain Research Center and the Goodman Faculty of Life Sciences, Bar-Ilan University, Ramat-Gan52900, Israel

*ido.kanter@biu.ac.il

The synchronization of chaotic lasers and the optical phase synchronization of light originating in multiple coupled lasers have both been extensively studied, however, the interplay between these two phenomena, especially at the network level is unexplored. Here we experimentally compare chaos synchronization of laser networks with heterogeneous coupling delay times to phase synchronization of similar networks. While chaotic lasers exhibit deterioration in synchronization as the network time delay heterogeneity increases, phase synchronization is found to be independent of heterogeneity. The experimental results are found to be in agreement with numerical simulations for semiconductor lasers.

Laser networks are a good experimental platform to analyze the interplay between network topology and synchronous mode activity of time delay coupled oscillators. The observed types of synchronization amongst two or more lasers can be divided into two types; optical phase synchronization of otherwise stable lasers and the synchronization of chaotic intensity fluctuations amongst optically coupled lasers. In the first type the intensity is practically constant and the optical phase is locked among the lasers [1-4], whereas in the second type the chaotic output intensity as well as the optical phase are the same, with zero time-lag, amongst all lasers [5-8]. There are geometrical configurations of networks, predicted theoretically and observed experimentally, where the network splits into several clusters, where lasers belonging to the same cluster are synchronized [9-11].

Experimental verification of chaotic intensity synchronization is limited to networks consisting of up to 4 lasers, as experimental synchronization of larger networks is difficult [12]. Hence, most of our knowledge of large network synchronization of chaotic lasers is based on simulations. Optical phase synchronization on the other hand, was recently experimentally demonstrated for a homogeneous time delayed coupled network having up to 16 lasers as well as for heterogeneous laser networks[13].

In this Letter we experimentally compare optical phase synchronization and chaotic intensity synchronization for a network of lasers with heterogeneous time delays. We find that chaotic intensity synchronization deteriorates as the time delay heterogeneity increases. In contrast, for the case of constant intensity phase synchronized lasers the optical phase synchronization is independent of the time delay heterogeneity. Given the limited experimental network size, such comparison may seem challenging. Here, we demonstrate a method that enables such comparisons even when using only two mutually coupled lasers and one self-feedback channel whose delay length is varied.

For *homogeneous* laser networks, where all coupling delay times are identical, the interplay between network topology and mode of synchronization is identical for phase and chaotic intensity [7, 11, 13]. Based on the theory of stochastic matrices, one can show that the mixing of information by the optical feedback leads to a number of synchronized clusters governed by the greatest common divisor (GCD) of the delay loops composing the network [14]. Hence, in a homogenous network with GCD=1 the lasers are synchronized isochronally, i.e. zero-lag synchronization emerges for both chaotic intensity and phase synchronization, provided that the coupling strengths are sufficient. Such a simple case is exemplified in Fig. 1(a) where the network consists of 2 lasers with $\tau$ and $2\tau$ loops and GCD(1,2)=1. Such universal behavior for homogenous networks was confirmed in simulations as well as experiments of chaotic intensity and optical phase synchronization [4, 6].

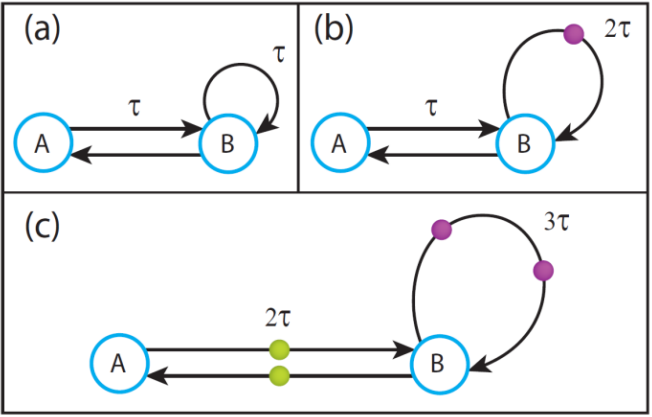

FIG. 1 (color online). Examples for the minimal number of additional imaginary lasers required to achieve a homogeneous network. Imaginary lasers on the mutual delay couplings, $\tau_c$, (green) and on the self-coupling delay, $\tau_d$, (purple). (a)$\tau_d/\tau_c$=1, 0 imaginary lasers. (b)$\tau_d/\tau_c$=2/1, 1 imaginary laser (c)$\tau_d/\tau_c$=3/2, 4 imaginary lasers.

For heterogeneous topologies, where self-feedback time delay differs from the mutual coupling delay (Fig. 1(b-c)), we define a quantity to measure the heterogeneity of the network. It is defined as the minimal number of additional imaginary lasers required to achieve an equivalent homogeneous topology. This quantity is exemplified in Fig. 1 for a network consisting of two lasers, A and B, with mutual delay time $\tau_c$ and a self-feedback delay, $\tau_d$, for laser B. In the case of $\tau_c=\tau_d$, corresponding to a homogeneous network, (Fig. 1(a)), no additional imaginary lasers are required. For the case $\tau_d=2\tau_c$ (Fig. 1(b)) one additional imaginary laser is required to divide the longer self-feedback delay into two equal $\tau$ delays and thus the heterogeneity equals 1. Figure 1(c) presents the case where $\tau_c=2\tau$ and $\tau_d=3\tau$, hence two imaginary (purple) lasers need to be inserted into the self-feedback delay and two additional lasers (green) into the mutual delays, giving a heterogeneity of 4.

In the following we first describe the two experimental setups: coupled semiconductor lasers synchronized in their chaotic intensity fluctuations and coupled solid state lasers with constant intensities, exhibiting optical phase synchronization. The later setup was already well examined in the literature [11, 13], however, without a systematic examination of the quantitative level of phase locking as a function of heterogeneity. Next we quantitatively show how the optical phase and chaotic intensity synchronization depend on the tenable global quantity - network homogeneity.

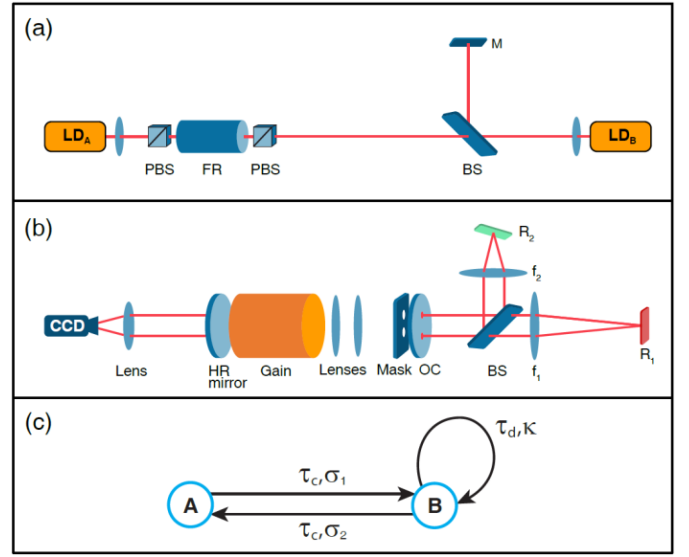

FIG. 2 (color online). (a) Chaotic synchronization experimental setup: mutually coupled chaotic lasers A and B with tunable coupling intensities and delay times. Laser B receives self-feedback from mirror M. Intensities are controlled using a Faraday rotator in conjunction with two rotated polarizing beam splitters (PBS). (b) Phase synchronization experimental setup: two mutually coupled independent solid state lasers. A degenerate cavity (middle), a coupling arrangement (right) and a detection arrangement used to detect phase synchronization between the two lasers (left). (c) Schematic representation of the two coupled lasers for (a) and (b): $\tau_c$, $\tau_d$ - mutual coupling and self-feedback delay times, respectively, $\sigma_1$, $\sigma_2$ - mutual coupling intensities of lasers B to A and A to B, respectively, $\kappa$- self feedback intensity for laser B.

The experimental setup for chaos synchronization is shown in Fig. 2(a). It consists of two similar Fabry-Perot semi-conductor lasers, emitting near 655 nm wavelength, and an arrangement for controlling the coupling and self-feedback intensities and delay times. Laser B receives delayed self-feedback from a mirror (M) in addition to the delayed coupling signal from laser A, via a coupling beam splitter (BS). To achieve synchronization, the total feedback intensity for each laser has to be similar, $\sigma_2+\kappa\sim\sigma_1$ (Fig. 2c) [6, 15]. Hence, to compensate the lower total feedback intensity for laser A, we set a different coupling strength for $\sigma_1$, from lasers B to A, than for $\sigma_2$, from A to B. This is achieved by means of a Faraday rotator positioned between two polarizing beam splitters (PBS) whose angles determine $\sigma_1$ and $\sigma_2$. The self-feedback time delay for laser B, $\tau_d$, and the mutual coupling time delay $\tau_c$ are tuned by controlling the distances of laser A and the mirror from the coupling beam splitter. Synchronization is measured by calculating the intensity correlation between the chaotic intensities of the two lasers [6]. Two fast photo-detectors biased via a bias T (not shown in the figure) are used to measure the laser's intensity. The AC components of the two laser intensities are measured simultaneously by two channels of a 12 GHz bandwidth, 40 GS/s digital oscilloscope (Tektronix TDS 6124C). Correlation coefficients are calculated between matching time segments of 10 ns length from each detector and then averaged over all segments from a total data stream of 2 µs length.

For the case $\tau_d=\tau_c$ (Fig. 1(a)), GCD(1,2)=1 and the isochronal intensity correlation is high, typically exceeding 0.9, (Fig. 3(a)). The intermittent drops in cross correlation, seen in Fig. 3(a), are a consequence of the well-known phenomenon of low frequency fluctuations (LFF), which occur when the lasers are operated close to their threshold current [16-17]. The synchronization of the chaotic fluctuations occur on a sub ns time scale as evidenced by the synchronization of the chaotic spikes which are typically of the order of 100 ps (Fig. 3(c)). For the case $\tau_d=2\tau_c$ (Fig. 1(b)), the network consists of identical delay loops of $2\tau$. Since GCD(2,2)=2, the two lasers belong to different clusters and the correlation at zero time delay vanishes (Fig. 3(b)), as shown in Fig. 3(d).

The experimental setup for phase synchronization is shown in Fig. 2(b). It consists of a degenerate laser cavity that can support many independent lasers [11, 18]. The figure shows the arrangement for controlling the coupling and self-feedback delay signal, and an arrangement for detecting phase synchronization by observing the interference fringes in the total laser output beam. The degenerate cavity is comprised of a Nd-Yag crystal gain medium that can support several independent laser channels, a flat 90% reflectivity (H.R.) front mirror and a rear 40% reflectivity output coupler (O.C.). Two lenses in a 4*f* telescope arrangement image the front mirror plane onto the rear O.C. plane whereby the transverse electric field distribution is imaged onto itself after propagating one full round trip of the cavity. Consequently, this allows any transverse electric field distribution to be an eigenmode of the degenerate cavity [18]. In particular, by placing inside the laser (adjacent to the O.C.) a mask with two holes of 0.4 mm diameter and 1 mm apart, two spatially localized and independent lasers emerge. Coupling between the two lasers is achieved by means of a coupling mirror, $R_1$ (in red) posited at the focal distance from focusing lens $f_1$, which is posited along one of the arms of a 50/50 beam splitter and set at a distaste of $f_1$ from the O.C. With $R_1$ aligned perpendicular to the optical axis each laser's output light is reflected back towards the other thereby mutually coupling them with a delay time of $\tau_c=4f_1/c$. In similar fashion, delayed self feedback is introduced by mirror $R_2$ (in green) and a focusing lens $f_2$ that are posited along the other arm of the 50/50 beam splitter. With the angular orientation of $R_2$ aligned appropriately, self feedback to only one of the lasers is achieved with a delay time of $\tau_d=4f_2/c$ [13]. The self and mutual coupling delay times were selected to be a few ns and were much longer than the coherence time of 10 ps for each individual laser, so the coupling signal arrives long after the phase memory is lost. The level of optical phase synchronization is quantified by calculating the visibility of the interference fringes measured in the far-field intensity pattern, as shown in Fig. 3(e-f) for $\tau_d=\tau_c$ and $\tau_d=2\tau_c$, respectively. Results clearly indicate that for $\tau_d=\tau_c$, GCD=1, the optical phase between the two lasers is synchronized (interference fringe contrast is ~0.8) (Fig. 3e), whereas for $\tau_d=2\tau_c$ (GCD=2) the contrast of the interference fringes visibility vanishes (Fig. 3(f)).

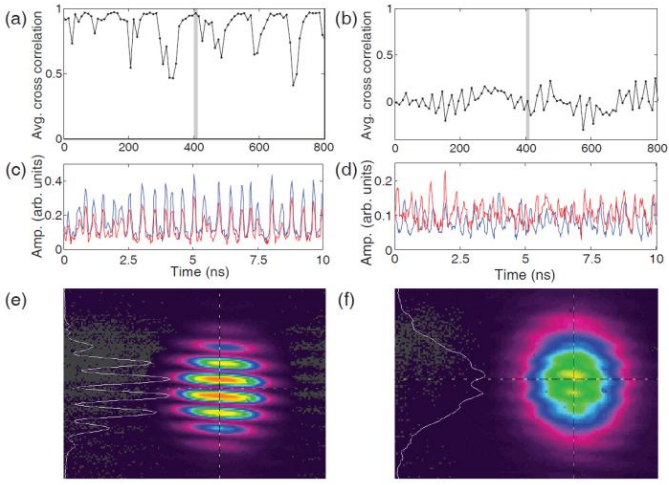

FIG. 3(color online). (a-b) Average cross correlation for chaotic lasers over 0.8μs recording for $\tau_d/\tau_c=1$ and $\tau_d/\tau_c=2/1$, respectively with laser injection current at 1.07 times threshold current. (c-d) 10 ns recording of the chaotic laser intensity corresponding to light gray area in (a) and (b). (e) Far-field intensity distribution for $\tau_d/\tau_c=1$ indicating zero-lag phase synchronization. (f) Far-field intensity distribution for $\tau_d/\tau_c=2/1$ indicating the lack of phase synchronization.

Figure 4 shows the quality of synchronization between lasers A and B as a function of the heterogeneity, which can be easily deduced from the top horizontal scale identifying the ratio $\tau_d/\tau_c$. The level of phase synchronization is quantified by calculating the visibility of the interference fringes measured in the far-field intensity pattern, as shown in Fig. 3(e-f). The level of chaotic amplitude synchronization is quantified by measuring the zero time lag correlation between the output intensities of two lasers [19]. Phase synchronization was examined with up to 14 imaginary lasers (red squares in Fig. 4). The results clearly indicate that as long as GCD=1, optical phase synchronization remains the same (~0.8), independent of the heterogeneity of the network. For GCD>1, e.g. $\tau_d/\tau_c=4/3$ resulting in two loops of $8\tau$ and $6\tau$ thus GCD=2, the optical phase synchronization is close to zero as expected [13].

The experimental results for GCD=1 were numerically confirmed using simulations of the Kuramoto model (with similar parameters as in [11]) which describes a general class of oscillators. They are in a good agreement with the experimental results indicating that optical phase synchronization is independent of the level of heterogeneity, i.e. the number of imaginary lasers (Fig. 4(a)). As opposed to phase synchronization, chaotic intensity synchronization exhibits a fast deterioration as the network heterogeneity increases. Figure 4 indicates that although GCD=1, chaos synchronization decays from a correlation ~0.86 for a homogeneous network ($\tau_d=\tau_c$) to 0.2 for a heterogeneous network with 8 imaginary lasers ($\tau_d/\tau_c=5/3$). The experimental results for the deterioration of chaotic intensity synchronization, as the network heterogeneity increases, were confirmed in numerical simulations using the Lang-Kobayashi rate equations [20]. The parameters used in the simulation are the same as in [6] with $P=I_{pump}/I_{threshold}=1.07$. For GCD>1, zero-lag chaotic intensity synchronization does not exist, Fig. 4(b), in either chaotic intensity or phase synchronization. Previously we have shown that intensity correlation is related to optical phase correlation and thus it is also expected to decrease with heterogeneity for the case of chaotic lasers [7].

The insensitivity of phase synchronization to network heterogeneity can be intuitively understood by the fact that each laser synchronizes to the delayed incoming signal and relays the optical phase information onwards in accordance to the network connectivity. Hence the propagation of the optical phase is the same with and without the imaginary lasers and the rule of the GCD and the level of phase synchronization is independent of the network heterogeneity. As opposed to phase relay, chaotic behavior incorporates some nonlinearity, and in particular the output intensity waveform of a chaotic laser differs from its input waveform, as a result of internal nonlinear processes of the laser cavity. Consequently, a heterogeneous chaotic network will demonstrate different dynamics than the equivalent homogeneous network with real lasers that replace the imaginary ones. This occurs as the self-consistent fixed point solution of zero-lag synchronization which is usually maintained for homogeneous chaotic networks, is now violated when heterogeneity is introduced. Intuitively this can be understood by considering a homogeneous network with GCD=1 where part of the lasers are replaced with imaginary lasers that only function as relays. As the number of replaced lasers increases, the perturbation from the chaos synchronization fixed point grows and deterioration in the correlation is expected.

The above intuitive explanation predicts that for a given heterogeneous laser network the weakening of chaos will result in the enhancement of chaos synchronization. Furthermore, in the limiting case where chaos disappears, a crossover to the behavior of optical phase synchronization is expected. This is indeed what we observe in simulations of chaos synchronization, exemplified for the setup of $\tau_d/\tau_c=3/2$ in Fig. 4c, where chaos is weakened by decreasing the pumping current towards the threshold P=1. As the pump to threshold current ratio, P, is decreased from P=1.07 towards P=1.0001, the correlation is enhanced from ~0.6 to ~1.0. Another way to weaken chaos is to decrease the coupling and feedback intensity to which the lasers are subjected. However, the correlation is expected to be non-monotonic in such a scenario as the lasers might pass strong-weak-strong chaos transitions [21].

In conclusion, we have addressed the question of how networks of coupled oscillators with self-feedback synchronize and how their zero time delay synchronization depends on network topology. Two types of synchronization were considered; in one the chaotic intensity fluctuations of coupled lasers can synchronize and in the other the optical phase of coupled lasers synchronizes. We have shown that for the case of phase synchronization network homogeneity is of little importance and the network synchronizes for any configuration for which the GCD is 1. For chaos synchronization, we also observe that GCD has to be 1 but here the heterogeneity of the network configuration has drastic effects on the synchronization which decreases with increasing heterogeneity. We expect the results for chaotic synchronization to be applicable to other chaotic systems, and the results for phase synchronization to be applicable for other models of excitable relaying units, such as spike activities in neural networks.

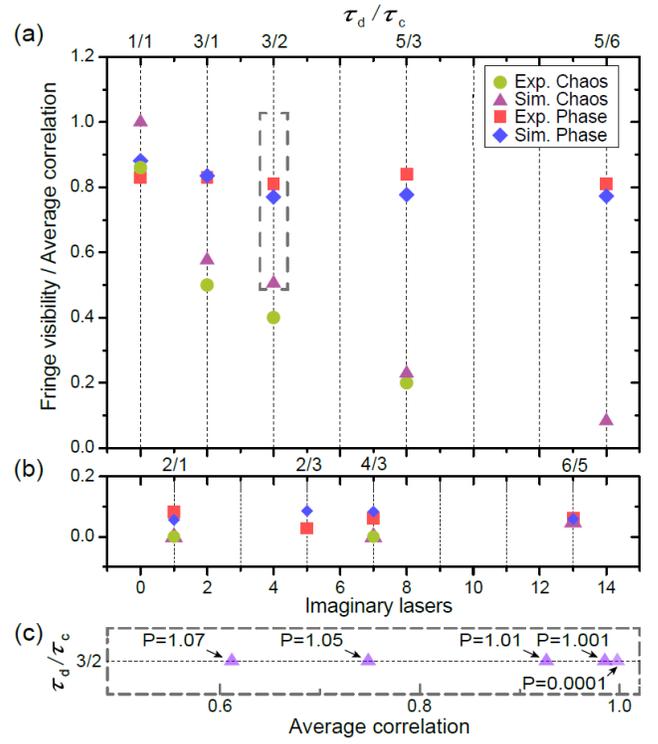

FIG. 4 (color online). (a) Average correlation and fringe visibility as a function of the heterogeneity as defined in the text, for topologies with GCD=1. Results for experimental chaotic intensity synchronization, (green circles) and numerical simulation (purple triangles), phase synchronization (red rectangles) and numerical simulation (blue square). Top horizontal scale identifies the self/mutual coupling delay times $\tau_d/\tau_c$. (b) Similar to panel (a) but for GCD=2. (c) Numerical simulation for the averaged correlation for $\tau_d/\tau_c=3/2$ for different injection currents, P=1.07, 1.05, 1.01, 1.001, 1.0001.

MR and IK acknowledge partial support from the US-Israel Bi-National Science Foundation.